**Efficient Choice, Inefficient Democracy?**

**The Implications of Cable and Internet Access for Political Knowledge**

**and Voter Turnout**

Markus Prior

Ph.D. Candidate
Department of Communication
Stanford University
McClatchy Hall, Bldg. 120
Stanford, CA 94305-2050
mprior@stanford.edu

**Efficient Choice, Inefficient Democracy?**

**The Implications of Cable and Internet Access for Political Knowledge**

**and Voter Turnout**

Abstract

This paper explains why, despite a marked increase in available political information on cable television and the Internet, citizens' levels of political knowledge have, at best, remained stagnant (Delli Carpini & Keeter, 1996). Since the availability of entertainment content has increased too, the effect of new media on knowledge and vote likelihood should be determined by people's relative preferences for entertainment and information. Access to new media should increase knowledge and vote likelihood among people who prefer news. At the same time, it is hypothesized to have a negative effect on knowledge and turnout for people who prefer entertainment content. Hypotheses are tested by building a measure of Relative Entertainment Preference (REP) from existing NES and Pew survey data. Results support the predicted interaction effect of media environment (cable and/or Internet access) and motivation (REP) on political knowledge and turnout. In particular, people who prefer entertainment to news *and have access to cable television and the Internet* are less knowledgeable and less likely to vote than any other group of people.

# Efficient Choice, Inefficient Democracy?
# The Implications of Cable and Internet Access for Political Knowledge and Voter Turnout

People face dramatically different media choices today than two or three decades ago. Television used to be broadcast only on three to five channels. Today, cable provides easily ten times as many channels and offers around-the-clock news coverage on several of them. The Internet further offers an unquantifiable amount of additional media options, including numerous newspapers, magazines, TV programs, and other political information. Few doubt that more information is available as a result of recent changes in the media environment. Does that imply that people are better informed? The present study attempts to answer this question by comparing people who have access to cable television and the Internet to those with access to only one of these new media and those who do not have access to either. Results suggest that some people with new media access may indeed be better informed than those with limited or no access. But other new media users actually know less about the political process than otherwise similar users of 'old' media. The knowledge gap (Tichenor, Donohue, & Olien, 1970) between the most informed and the least informed is larger among new media users than among people without access to cable or Internet. This larger gap, I argue, is the result of the parallel increase of news *and entertainment* options for new media users. Cable viewers and Internet users can watch, read or listen to abundant information, but they can also avoid news better than people without the Internet and only broadcast television. The challenge is to predict who will indulge in news and who will ignore it.

In a broadcast environment, audiences are considered "captive". When referring to broadcasts of presidential addresses, Baum and Kernell (1999, 101) maintain that "a viewer's 'captive' status results from the combination of limited channels, an unwillingness to turn off the set, and the networks' joint suspension of commercial programming during a presidential appearance." While presidential appearances broadcast simultaneously on all networks represent an extreme case, broadcast viewers face a comparable situation every day in the early evening, when most, if not all, broadcast channels offer local and national news for at least an hour. At certain times of the day, network programming strategies effectively reduce opportunities for broadcast viewers to a choice between watching news and turning off the TV. Cable and Internet access remove these constraints by making more content and content types available overall and at any point in time. For people with access to new media, content preferences should determine exposure more directly than for the still-captive viewer without cable or Internet access.

Content preferences condition the effects of media use, and the power of this conditional effect increases with the number of media choices. Baum (1999; Baum & Kernell, 1999) introduces a model of television watching, in which people's expectation about the utility they can gain from different programs determines their program choice. People weigh the benefits of obtaining information against the transaction costs of obtaining the information and the opportunity costs of not being able to use their attention differently. Baum argues that people did not tune in to hard news programs because opportunity costs (in form of forfeiting payoffs from entertainment) were too high. With more soft news programs, this utility calculation changes and more people maximize their utility by watching soft news (i.e., a mix of news and entertainment), because it provides a sufficient amount of entertainment. This paper applies a similar logic to the more general (and simplified) choice between news and entertainment content, focusing not only on changes due to increased program availability on cable television, but also the impact of the Internet.

One of the conclusions from Baum's model is that exposure to news and political awareness of certain events is not always the intended consequence of people viewing decisions, but can also occur as a "byproduct". News-watching is not only determined by people's desire for information, but also by the entertainment programs the they miss while watching news. The amount of news people watch is in part a function of the availability of non-political programs and people's liking of these non-political programs, compared to news. Since news exposure leads to learning (Neuman, Just, & Crigler, 1992), political knowledge can be a byproduct of a more general 'utility' calculation not intentionally or exclusively focused on obtaining political information.

In a model of viewing decisions similar to Baum's, Becker and Schönbach (1989) combine uses-and-gratifications and expectancy-value theory to emphasize the difference between the primary effect of satisfying needs through media use and secondary effects such as knowledge gains. Recent work by Putnam and his colleagues (Campbell, Yonish, & Putnam, 1999; Putnam, 1995, 2000) focuses on civic engagement as a specific secondary effect of television-watching, emphasizing different consequences of exposure to news and entertainment programs. Participation increases with news consumption, but decreases with exposure to entertainment programming. While concerns about possible reverse causation warrant some caution in interpreting these results, they do emphasize that different program types, and, by extension, different program preferences may have different political consequences.

Consistent with Becker and Schönbach (1989) and Baum (1999), Putnam's work shows that political consequences are often indirect (secondary), rather than direct and actively sought. I argue that the possibility for cable viewers and Internet users to better match content preferences and content choices reduces situations in which political knowledge is obtained as a secondary consequence or "byproduct." In this paper, I develop hypotheses about the effects of new media on people's political knowledge and likelihood to vote, based on the premise that content preference and content choice match more closely for new media users. The hypotheses are tested using existing survey data. Despite difficulties in creating a measure of motivation from available data, empirical tests support the hypothesis that content preference is a better predictor of political knowledge and vote likelihood for new media users than it is for people with limited or no access to new media. People who enjoy watching entertainment more than news *and have access to* *cable television and the Internet* are less knowledgeable and less likely to vote than any other group of people.

**Political Learning in a Media Environment with Increased Choice**

Numerous studies, most of them in the uses-and-gratifications tradition (Katz, Blumler, & Gurevitch, 1973; Katz, Gurevitch, & Haas, 1973), have shown that people use media for different reasons. One of the most basic distinctions is between entertainment-focused and information-focused media use (e.g., Rubin, 1984). Most entertainment exposure is motivated by the expectation of immediate, diversionary gratification. People do watch entertainment programs on television to gain knowledge about social behavior or fashion, among other things, but "[t]hese guidance motives are generally moderate in importance, ranking below enjoyment-oriented reasons" (Atkin, 1985, 87,88). According to one recent analysis of survey data, more people give as their reason for watching television "to be entertained" than "to learn something" or "to keep up with what's going on." For roughly forty percent, being entertained is the primary reason for watching. Another fifty percent mention both being entertained and informed (Campbell et al., 1999). Similar results were obtained by other studies (e.g., Comstock & Scharrer, 1999; Graber, 2001).

Yet while a sizable segment of the population watches television primarily to be entertained, and not to obtain political information, this does not necessarily imply that this segment is not also exposed to news. When only broadcast television is available, the audience is captive and, to a certain extent, watches whatever is offered on the few television channels. Audience research has confirmed a two-stage model according to which people first decide to watch television and then pick the available program they like best. Klein (1972, 77) aptly called this model of television the "Theory of Least Objectionable Program". Empirical evidence for the two-step model comes from analysis of audience data showing repeat-viewing rates of around 50% (Barwise, Ehrenberg, & Goodhardt, 1982). That is, only about half the viewers of a particular program watch the same program on the following day. Repeat-viewing rates are low for all program types and do not increases when repeat-viewing of genres instead of particular programs is evaluated (Barwise & Ehrenberg, 1988, 40). During the heydays of broadcast television, one study showed that only a third of all programs are watched from beginning to end by at least 80% of the people watching the programs at some point. 40% of the respondents reported watching programs because it came on the channel they were already watching or because someone else wanted to see it (LoSciuto, 1972). Hence, a major share of television viewing does not seem to follow a deliberate choice of a program, but convenience, availability of spare time and the decision to spend that time watching television (see also Comstock & Scharrer, 1999; Neuman, 1991, 94).

It follows from the "two-stage" viewing behavior that news audiences should be larger when no (or few) alternatives are offered on other channels. Indeed, local news audiences tend to be larger when no competing entertainment programming is scheduled (Webster, 1984; Webster & Newton, 1988; Webster & Wakshlag, 1983). Based on the analysis of commercial audience data, Barwise and Ehrenberg (1988, 57) conclude that restriction of choice due to simultaneous scheduling of news programs on all networks "goes with lower audience appreciation among those viewing at these times (i.e. *they did not all necessarily want to watch news then*)" (emphasis added). As cable viewers can easily evade simultaneous news programs on all or most broadcast channels, audiences for entertainment programs should increase at the expense of news among people with cable access. Weimann (1996) documents that the introduction of cable in Israel caused increased watching of movies and MTV-like music channels (see also Katz, 1996). Correlational research for the United States suggests that cable subscribers watch more entertainment programs than non-subscribers (Becker, Creedon, Blood, & Fredin, 1989). The most convincing evidence that high news ratings among broadcast viewers are explained by lack of alternatives rather than preference comes from a study by Baum and Kernell (1999) who show that cable subscribers, especially the less educated among them, are less likely to watch the presidential debates than people who receive only broadcast television. Far from being conclusive, these studies do suggest that viewers without cable access are constrained by the limited opportunities to watch entertainment programs and that they would prefer to tune in to entertainment more frequently than the offerings on broadcast channels allows them to do. And hence, "[a]lthough cable has fostered a core of "news junkies" who immerse themselves in CNN and C-SPAN, its more significant effect has been to contribute to a steep decline in the overall size of the news audience." (Patterson, 2000, 247)

Once exposed to television news, people learn about politics (e.g., Neuman et al., 1992; Zhao & Bleske, 1995; Zhao & Chaffee, 1995). Exposure alone leads to learning; attention is not necessary to pick up at least basic facts from the news (Keeter & Wilson, 1986; Zukin & Snyder, 1984). Zukin and Snyder (1984) show that even many politically uninterested New Jersey citizens who received their broadcast news from New York City stations recalled the names of New York's mayoral candidates, even though they could not vote for any of the candidates.[1] Hence, broadcast viewers are likely to learn about politics even in the absence of political interest. Even those who would prefer to watch entertainment programs rather than news acquire at least basic political knowledge more or less "accidentally," because they happened to turn on their television at a time when only news was on.

This kind of accidental exposure and passive learning is much less likely among cable viewers. With plenty of entertainment options at all times of the day, cable viewers do not have to settle for news as their "least objectionable" program choice. Hence, for viewers who prefer entertainment to news, political knowledge should be lower if they have cable access than if they can only watch over-the-air channels. Since, on the other hand, cable viewers with a preference for news have the opportunity to watch more news than broadcast viewers with the same preference, political knowledge should be higher among cable viewers with a news preference than among otherwise similar viewers without cable access. In short, theoretical considerations lead me to predict an interaction effect between cable access and content preference on political knowledge.

---

[1] One of the earliest references on the subject of "passive learning" is Krugman and Hartley (1970) who discuss learning from television among children.

Much of this section focused on television viewing behavior since decades of research have produced relatively firm understanding of what people like about watching television and how they decide what to watch. Research on Internet use has not yet developed as far, in part because the medium itself changed quickly in its first decade. Consequently, predictions about the effects of Internet access cannot be based on an equally developed theoretical understanding of user behavior. With respect to one key property, however, the Internet is very similar to cable television: Both increase the availability of media content considerably. To the extent that greater availability and choice alone explain the effect on political knowledge derived above, the effect of the Internet may be very similar to that of cable television. Chance encounters with political information (beyond the occasional headline) may be equally infrequent on the Internet. Active searching for political information driven by a preference for news may be required for people to learn about politics. Avoiding politics may be equally easy (if not easier) on the Internet as on cable television. The following analysis should be understood as a first test of the proposition that increased content availability and choice on cable TV and the Internet have similar effects, *not* as an argument that cable television and Internet are equivalent *in all respects*.

In the second part of this paper, the analysis is extended to turnout. Since an individual's likelihood to vote in an election increases with her level of political knowledge (Tan, 1980; Verba, Burns, & Schlozman, 1997; Verba, Schlozman, & Brady, 1995, ch.12), the same interaction between new media access and content preferences should occur for models of turnout. All other things equal, new media users with a preference for news should be more likely to vote in an election than people with the same news preferences, but limited or no new media access. Analogously, people with access to new media and a preference for entertainment should be less likely to vote compared to otherwise similar individuals without new media access.

**The Data**

Survey data to test the model must include information about respondents' cable and Internet access, their relative entertainment preferences, and at least some measures of political information and behavior. This turns out to be a rare combination. The 1996 and 2000 National Election Studies include the items on political information and voting in presidential and House elections, in addition to at least elementary items on entertainment and questions on cable and Internet access. As for Pew data, the Media Consumption Surveys conducted in 1996, 1998, and 2000 include questions on voting and entertainment items that allow an alternative operationalization to that based on the National Election Studies. Testing the present model is severely limited by the available survey data, but not impossible.

*Relative Entertainment Preference*

Testing the hypotheses requires a measure of people's relative preferences for entertainment over news. Since respondents are not directly asked about their preferences in any of these surveys, preferences have to be assessed indirectly by the actual program types they report watching. In particular, relative entertainment preference is measured as the share of entertainment viewing, or:

$$\text{Relative Entertainment Preference} = \frac{\text{Entertainment Viewing}}{\text{Entertainment Viewing} + \text{News Viewing}}$$

Entertainment viewing is measured slightly differently for each of the surveys used here. While consistency was the goal in creating the scales, the available items differ from survey to survey. For the NES 2000, entertainment viewing is based on two items about *Jeopardy* and *Wheel of Fortune*. Entertainment viewing is measured by the mean daily viewing of both shows. For the NES 1996, entertainment viewing is the mean frequencies of watching games shows ("*Jeopardy* or *Wheel of Fortune*") and *Dr. Quinn, Medicine Woman* (measured on a four-point scale and standardized to the 0-1 interval). For the Pew Media Consumption Surveys, entertainment viewing is operationalized as watching *Entertainment Tonight* (MCS 2000),

*Entertainment Tonight* and MTV (MCS 1998), and *Hardcopy* and MTV (MCS 1996). All items used four-point response formats ("regularly", "sometimes", "hardly ever", "never"). The Pew studies did not include questions about pure entertainment options other than the MTV item.

News viewing is operationalized as average daily viewing of local and national news for all data sets.[2] REP cannot be computed for respondents that reported no information and entertainment viewing at all. These respondents are excluded from the analysis (2.5% in MCS 1996, 2.8% in MCS 1998, 5.4% in MCS 2000, 5% in NES 1996, 11% in NES 2000). A summary of the relative entertainment preference scales and descriptive statistics is in appendix table A.1.

Obviously, the entertainment programs on which the scales are based are arbitrary. Arguably, game shows are not a viable entertainment options for many people, not even in times of severe boredom. Given that this study is limited by the availability of secondary data, there is no fix for this problem. Using a number of different entertainment programs in different surveys, however, should ease fears that results are based only on the idiosyncrasies of game show viewers. In order to assess the seriousness of these inevitable problems, the next section provides a validity check for the measures of relative entertainment preferences.

*Validating the Index of Relative Entertainment Preference*

The basic premise of this paper is that people, if they have the opportunity, expose themselves to media content that they like. This idea is simplified here to distinguish people who prefer news from those who prefer entertainment content. Since no direct measure of content preferences is available, the index of relative entertainment preference (REP) was created as a ratio of self-reported exposure to news and entertainment (news) programs. In this section, I present some evidence that the REP index, while created differently for different data sets, is consistently and in intuitive ways related to other relevant variables.

---

[2] The 2000 NES asked separately about early and late local news. Only the frequency of watching early local news is used here.

Table A.1. shows bivariate correlations of REP with various measures of attention to news. For the NES 1996 and 2000, REP is negatively related to local and national news attention. Consistent with the notion that national news tend to be more serious, the bivariate correlation is slightly larger for attention to national news. For the Pew data sets, table A.1. contains bivariate correlations of the REP index with respondents' self-reported tendency to follow general news and entertainment news in particular. For all three data sets, REP is negatively related to following news. More importantly, however, the correlations of the REP index with following entertainment news are distinctly more positive, indicating that the REP index distinguishes people with a preferences for hard news from those with a preference for entertainment aspects of news, if not entertainment per se. (The Pew studies do not include any measure of viewing entertainment shows, so the relation of REP and entertainment programming cannot be assessed.)

The second way to validate the REP index is to examine its relation to television news exposure. Theoretically, news exposure is the mediating variable between REP and political knowledge. The prediction that the knowledge gap between information- and entertainment-seekers is larger for cable subscribers is based on the intuition that information-seekers can watch more news when they have cable access, while entertainment-seekers with cable can avoid news more effectively. Thus, the interaction of cable and REP should be negative in its effect on news exposure.

Table A.2. presents empirical tests of the prediction for various measures of television news exposure. In order to be useful for validation purposes, the items have to be general enough not to mention a particular news program or even "network news," as such measures would have different meanings for people with and without cable access. (A person without cable access reporting to watch little network news is unlikely to be exposed to a lot of national news, whereas this conclusion would be invalid for someone with access to cable.) Pew's 1996 MCS includes a 5-point scale of time spent watching television news on the day before the interview[3], as well as the question whether the respondent watches more news than before.[4] In the 2000 MCS, respondents were simply asked whether they watch news regularly or not. For the NES 1996 and 2000, finally, composite news exposure measures were built from self-reported watching of the presidential debates and "programs about the campaign" (alpha=.68, r=.52 for NES 2000; alpha=.69, r=.52 for NES 1996).

Results in table A.2. are very consistent. Exposure to television news is lowest among people with cable access and a strong preference for entertainment. Access to cable television increases the effect of entertainment preference on news viewing. This interaction is significant at p<.10 or better in four of the five models and has a p-value of .19 for the fifth model.[5] The consistently very low $R^2$s, however, should caution against overinterpreting these results. They do provide indication that the REP index indeed measures the concept of relative content preferences.

*Information and Participation Measures*

The dependent variables in this paper are a set of information measures about congressional elections and voting in House and presidential elections. The NES 1996 and 2000 include the traditional questions on name recall for House candidates in the respondent's district. The number of correctly recalled candidate names is counted so that the measure ranges from zero (no recall) to two (recall of both candidates' names). In addition, indices of incumbent-specific information are created in the NES data sets from four items. In 1996, respondents were asked whether the

---

[3] The same measure could be built for only one half of the sample in the MCS 1998 and 2000. Results were not significant and are not reported here.

[4] This index is built from two three-point scales asking the respondent to indicate whether she watches more, less, or the same amount of local and national news.

[5] In addition, table A.2. suggests a positive effect of entertainment preference on news viewing for respondents without cable. The effect is significant and sizable in only two of the five models, but does indicate the possibility that the REP index picks up the amount of TV consumption in addition to the relative preference for entertainment.

There is some evidence for this from the two other NES pilot studies that contain measures of overall viewing (but not data on internet access). The correlations between REP and total viewing at night and during the day are r=.17/r=.16 for 1995 and r=.09/r=.09 for 1998) For the purpose of testing my model, this possibility is not a severe problem since, to the extent that high REP is correlated with watching a lot of television, the estimated effect of REP will be biased downwards.

incumbent voted for or against welfare reform, how often the incumbent supported President Clinton's legislative proposals, and how well the incumbent "keeps in touch with the people in your district". The number of times that the respondent gave an answer other than "don't know" was counted. The forth item was the respondent's knowledge of who the incumbent candidate in her district was. For the 2000 index, the first two items were different: Respondents were asked whether they remembered anything the incumbent had done for their district and whether they knew the number of years the incumbent was in office.

Vote measures were created for self-reported vote in presidential and House elections (post-election interview) for the 1996 NES.[6] The Pew Media Consumption Surveys in 1996, 1998, and 2000 use retrospective questions on voting in the last presidential election. Any specific candidate named is coded as 1; "did not vote," "don't know," and not remembering which candidate the vote was cast for are coded as 0.[7]

*Cable Access and Control Variables*

Cable access is coded as a dummy variable based on the question "Do you have either cable or satellite television?" (Hence, "cable access" subsumes satellite services.) To test whether the

---

[6] Unlike before, one half of the 2000 NES interviews were conducted by telephone. Interview mode affects responses. For example, 76% of the respondents interviewed by phone reported voting in the 2000 presidential election, while only 68% of the people interviewed face-to-face did. The different degrees of overreporting as a result of different interview modes affect the estimation of the vote model for the NES 2000 considerably. In fact, the coefficients for the interaction effect are opposite for phone and face-to-face interviews, and the explained variance is lower by a third in the phone condition. For the remaining half-sample interviewed face-to-face, the effect is in the predicted direction, but does not reach significance . Since it is unclear how to treat the different interview modes, I decided not to include the NES 2000 in the analysis of turnout effects.

[7] Respondents who did not know whether they voted or who refused to answer (1996: 3.8%, 1998: 6.1%, 2000: 7.3%) could be excluded from the analysis. Respondents who reported voting, but did not remember whom they voted for (1996: 5.1%, 1998: 6.7%, 2000: 8.8%), could be coded as not having voted. The results change very little for these alternative specifications for the 1996 and 1998 MCS.

In the 2000 MCS, Pew used a modified coding scheme, reporting mention of other candidates than Clinton, Dole or Perot. This new category is coded as having voted. The analysis for the 2000 MCS turns out to be somewhat sensitive to the coding of respondents that reported voting but do not remember the candidate they voted for. For the 2000 MCS only, results are reported for coding these respondents as having voted because it improves the fit of the regression as measured by the explained variance. While coding them as not having voted (as in the MCS 1996 and 1998) reduces the size of the interaction coefficient and the significance level (from $p<.01$ to $p<.10$), the hypothesis is supported either way. Estimations of all models are available from the author.

effect of cable and the effect of new media more generally are similar, a measure of new media access is created. Respondents with access to either cable television or the Internet are scored 1, while respondents with access to both are scored 2. People without access to either receive a score of 0.

Two possible confounding factors have to be considered carefully in the present study design. First, new media access as an explanatory factor has to be distinguished from the influence of other variables that might explain information and turnout, and *that are correlated with having new media access*. Most importantly, cable and Internet access is not affordable for all Americans. This leads to correlations between cable access and other demographic variables, notably income, education and campaign interest. While a cross-sectional design is certainly not optimal to disentangle these influences, the present study uses several demographic control variables, including income, education, and campaign interest. They are described in detail in the appendix.

The second possible confound results from the self-selection of cable subscribers and people with Internet connections. Some people might be more interested in political affairs (and already more knowledgeable) and obtain new media access only as a result of their existing higher interest. This concern is eased by including a variety of variables that control for possible difference between cable subscribers and broadcast viewers. These control variables include campaign interest, political information, frequency of discussing politics, group membership, trust, internal and external efficacy, and the strength of the respondent's partisan identification. Hence, the analysis compares the effect of new media access among otherwise (in a statistical sense) similar individuals—similarly interested in, and informed about, politics; similarly trusting in government; and similarly engaged in the political process.

Moreover, differences in knowledge could be explained simply by different media use patterns that happen to correlate with new media access. In order to minimize this possibility, controls are included for respondent's frequency of watching national and local news, reading the newspaper, and listening to talk radio. Since different media uses tend to be positively correlated, this also controls for the possibility that people with new media access simply spend more time using media and are better informed as a result of their long watching or web-surfing hours, not their cable or Internet access as such. To be even more certain that the length of media use does not confound the analysis, a control variable is included for the time that respondents spent working. These and additional control variables are described in detail in the appendix. Some were not available for all data sets used in this study. Using a variety of control variables reduces the danger of attributing effects to new media access and entertainment preferences that are in fact caused by correlates of those factors.

Most of the control variables are almost completely observed, but a considerable number of respondents refused to answer questions about their income. On average about 15% of the respondents have missing data on this variable in the data sets used here. Since cable and Internet access may not be affordable to respondents with lower income, including income as a control variable is important. The failure to do so might lead to the incorrect attribution of knowledge differences to new media access, when in fact income differences are the causal factor. Excluding all respondents with missing value on the income variable, however, introduces a different bias if refusing to answer questions on income is systematically related to other characteristics of the respondent. A solution to this problem is to impute values for respondents that failed to answer a question. Based on other information about the respondent, the best estimate of their income (and a measure of the uncertainty of this estimate) is created. This is done via multiple imputation (King, Honaker, Joseph, & Scheve, 2001).

For this analysis, missing data on all independent variables except for the REP index and access to cable and new media are replaced by imputed values using the EMis algorithm implemented in AMELIA (Honaker, Joseph, King, Scheve, & Singh, 1999). Missing values on REP, new media access or any of the dependent variables are not imputed since their relationship to other variables in the data sets that could be used for imputation is unclear. When missingness is "nonignorable" (NI), that is, a variable with missing data is not related to any other variable in the data set, listwise deletion outperforms multiple imputation. While this cannot be directly tested with only the observed data, not imputing missing values on REP, new media access, and the dependent variables seems the more conservative approach. (New media access and the dependent variables have very little missing data anyway.) The following tables give the number of observations included in the analyses with and without multiple imputation. Using multiple imputation increases the number of cases by between 10 and 20% and helps to avoid bias from non-random deletion of cases.

**Results**

According to the first prediction, viewing preferences should be a better predictor of political knowledge for people with greater access to new media. Consequently, the interaction effect of REP and new media access in regressions of the different information measures should be negative and significant, whereas the main effect of REP should be insignificant. In particular, knowledge should be lowest for people with maximum access to new media and a strong preference for entertainment, as these people are most likely to use new media to avoid news exposure. In the following analysis, the predicted values on the knowledge scales obtained from the regression models should be lowest for this group.

This section presents the results both for the effect of cable television only and for the effect of the new media. For recall of House candidates, results from OLS regressions are shown in table 1. The interaction term is significant at *p*<.10 in three of four models, and marginally significant at *p*=.14 in the fourth. To interpret the results correctly, main and interaction effects need to be considered together. Figure 1 does this by graphing the predicted values by new media access for the range of possible values that relative entertainment preference can take, while holding all other independent variables at their means. (Only results from the new media models are graphed.) The models include controls for a variety of factors that would otherwise confound the impact of new media access and entertainment preference. Most importantly, respondents' overall political knowledge, as measured by interviewer assessment, and campaign interest are held constant. In other words, candidate knowledge is lower among respondents with greater new media access and a preference for entertainment than among equally informed and interested respondents with less access. Controls are also included for respondents' media use habits and frequency of political discussions. That is, people know less about the candidates when they have an entertainment preference and new media access, even if they report watching the same amount of local and national news, reading newspapers with a similar frequency, listening to talk radio for equally long periods, and discussing politics at similar frequencies. Finally, the models control for factors such as age, gender, income, and working hours that might affect media use and opportunities.

(Table 1 about here)

The graphs in the first row of figure 1 provide support for the hypothesis: Among people who prefer entertainment, greater access to new media is associated with lower knowledge about politics. The reverse is not apparent in the results: People with a preference for news and access to new media do not recall candidate names better than people with a news preference but no or only limited access to new media. One graph (NES 1996) also suggest a slight *increase* in recall and familiarity at higher levels of entertainment preference for people without new media access. This main effect of the REP index is positive in some of the models, sometimes at statistically significant levels. One explanation is that the measure of entertainment preference picks up the amount of media use to some degree, as shown above.[8] With respect to cable television, it is not inconceivable that political advertising, which is often most heavily targeted towards prime-time (i.e., non-cable) programs, affects broadcast viewers with high relative entertainment preference disproportionately. Given the limitations of the available measures, it is difficult to interpret the absolute predicted values. Their relative magnitude, however, shows that a high preference for entertainment translates into below-average political knowledge *only for people with access to new media.*

(Figure 1 about here)

The second measure of political knowledge employed here focuses on incumbent-specific information. This measure includes information about who the incumbent in the race is and about her record while in office. Results for 1996 and 2000 are presented in table 2. The interaction effect is in the predicted (negative) direction and significant at $p<.05$ for three of four models. The specification that includes both cable and Internet in measuring new media access performs better than the cable only model. The second row of graphs in figure 1 illustrates results graphically by showing predicted knowledge levels, holding other variables at their means. Again, knowledge decreases with REP only for people with new media access, and more so if they have access to both cable and Internet. However, as for recall, there is no discernable effect for low REP and the main effect of REP is again positive for the NES 1996 data. In sum, the analysis can only support a weaker version of the hypothesis for political knowledge: The more people have access to new media, the less they know about (congressional) politics, if they prefer entertainment. The reverse does not find support: Among people who like news, access to new media does not appear to increase political knowledge.

(Table 2 about here)

Since politically knowledgeable people are more likely to vote (e.g., Verba et al., 1997; Verba et al., 1995, ch.12), viewing preferences should also be a better predictor of turnout for those with greater new media access. If, as results suggests, people with a strong entertainment preference and maximum new media access are indeed the least knowledgeable segment of the electorate, they may also be the least likely to vote. Table 3 presents tests of this second prediction for voting in the 1996 presidential and House elections based on data from the NES 1996. Support for the hypothesis comes from voting in the House election. Figure 2 graphs the likelihood of voting by new media access for the range of the entertainment preference variable holding other variables at their means. The shape of the interaction for the 1996 NES Study, while in the predicted direction, is somewhat unexpected, since it appears to be driven mostly by changes among viewers without new media access. As in the discussion of the information results, the relative differences are more insightful than the absolute values. Among people with a strong relative entertainment preference, the likelihood to vote decreases with access to new media.

---

[8] See footnote 5.

(Table 3 about here)

(Figure 2 about here)

As table 3 also shows that the hypothesis is not supported for voting in the 1996 presidential election, results from the NES 1996 are mixed. In order to provide additional tests, models of vote likelihood are estimated on data from the Pew Media Consumption Surveys conducted in 1996, 1998, and 2000. Tables 4 and 5 present results for logistic regression models of voting in the last presidential or House election before the survey based on the Pew data. Graphic illustrations of the coefficients are in figure 2. Of the eight models, six yield significant interactions at $p<.10$ or better, while the coefficients in the two other models approach significance at $p=.12$ and $p=.15$. These results provide the clearest support for the hypothesis. Entertainment preference affects the likelihood to vote very little among people without any access to new media. Among people with access to cable and Internet, on the other hand, the REP index is strongly related to vote likelihood.

(Table 4 about here)

(Table 5 about here)

Moreover, the effect among people with a high news preference is more symmetric than for the models of political knowledge. While news-seekers with new media access were not discernibly more knowledgeable than those without, people with the same high preference for news are indeed more likely to vote when they have access to new media. As in the knowledge models, the reverse is true for people with a preference for entertainment. While a drop in knowledge and turnout among people with high REP and access to new media is clearly apparent throughout this analysis, more empirical tests are required to establish whether the additional news offerings on cable and Internet have a positive effect on these variables among people with a preference for news.

New media access, then, appears to increase the effect of relative entertainment preference on political knowledge and vote likelihood. But how certain can we be that this reflects the increased media choice for cable subscribers and Internet users, not some other characteristic that they share? One way to guard against such a confound is the use of control variables. But the controls used in this study cannot entirely rule out that the technological savvy of new media users or their 'lifestyle' accounts for the observed results. To examine this claim, I made use of a number of variables in the MCS 2000 that asked respondents whether they owned a cell phone, a pager, a DVD player, and a palm pilot. The same model of vote likelihood as in table 5 was run with these variables instead of cable access. While the coefficient for the interaction with new media access was significant at $p<.01$, none of the interactions of the REP index with the other technological devices produce results that even approach significance. Consequently, increased media choice, not some other aspect related to cable technology or its users, is likely to cause the observed effects.

It would be possible that the results presented here simply showed that more politically interested people watch more news and become more informed if they have access to new media. Entertainment, in this view, does not matter at all and the results occur regardless of the role of entertainment content. I tested this counter-hypothesis in two ways. If the above were true, running the analyses with an interaction of new media and political interest instead of the REP index should produce comparable, if not stronger, results. However, interaction terms of new media access with political interest (or education) do not reach statistical significance. Second, if entertainment preferences did not have any effect on knowledge or turnout, the effect of the REP index should be similar to the effect of news viewing alone. However, including an interaction of new media with news viewing instead of the complete REP index did not produce significant estimates in any of the models. The negative results indicate that political interest by itself does not explain the effect of new media on knowledge and vote likelihood. Rather, entertainment as a competing choice, its relatively greater availability, and its appeal to some media users drive the observed results *in conjunction with the motivation to follow politics*. The relative balance between preferences for these two broad types of content, news and entertainment, explains people's viewing decisions better than either one by itself.

This result was to be expected given that political interest and entertainment preference are largely independent. A person who reports to be highly interested in politics does not necessarily watch or read a lot of news, because the same person could also be very attracted to entertainment content. Similarly, if you are not very interested in politics, you may still be exposed to news if entertainment is even more boring to you. That neither the new media × political interest interaction nor the new media × news viewing interaction produces significant results, while the new media × REP interaction does, suggests that new media users with a preference for entertainment know less about politics because they like entertainment, not because they are uninterested in politics.

**Conclusion**

Andrew Kohut (quoted in Marks, 2000), the director of the Pew Research Center, recently asserted that "[c]able is the political conduit of the air. If you don't have that cable coming in your house, you're getting a whole heck of a lot less information about politics." This study suggests that having "that cable" or new media in general "coming in your house" does not automatically make people better informed, and, in fact, may lower political knowledge for some of them. The simple reason is that new media bring the opportunity to learn about politics at every hour of the day, but also offer the chance to avoid politics entirely:

> The period since the late 1980s is often considered the beginning of the "information age." (…) What has been mostly overlooked is that the indicated technological development toward increasingly rich media environments carries with it a previously unimaginable wealth of entertainment choices. In fact, entertainment offerings obtrusively dominate media content and are bound to do so in the foreseeable future. This circumstance, together with the apparent growing public demand on entertainment provisions, land equal justification to characterizing the present times as the "entertainment age." (Zillmann & Vorderer, 2000, vii).

The current period may be information age for some; for others, it is the entertainment age. This study demonstrates that we need to know people's content preferences in order to assess the political implications of the information/entertainment age. I have tried to measure this preference through an index of relative entertainment preference (REP). Built from existing data not collected with this goal in mind, the REP has several weaknesses. It is based on self-reported exposure to entertainment and information programs. The selection of these programs is necessarily arbitrary and severely restricted by the available data. Ideally, an index of REP would not rely on exposure measures at all, but ask respondents about their relative preference for different types of content.

The operationalization of REP has some empirically undesirable qualities. Since few people report watching a lot of entertainment programs but little or no news (and since people tend to exaggerate their news consumption), the index is right-skewed. Moreover, it is slightly correlated with the overall amount of television viewing. Therefore, the interpretation of the absolute predicted values obtained from the models estimated here cannot be entirely conclusive. Some models, notably of political knowledge, indicate a trend towards slightly increased knowledge among people with a preference for entertainment and no access to new media. Future research will have to create a more reliable measure of REP to determine whether this trend is a measurement artifact or an observation that is stable across measures.

Despite these weaknesses, the results in this study provide consistent empirical support for the hypothesized processes. In an attempt to compensate for the inherent arbitrariness of the operationalizations, a number of different data sets are used to test the hypotheses. Data collected by different organizations (NES and Pew) and for different election contexts support the hypotheses consistently and at acceptable levels of statistical significance given the imprecision of the main independent variable. Moreover, while caution is warranted when interpreting the absolute predicted values, confidence in the relative effects of new media access and entertainment preference is much higher. Throughout the analysis, people with a preference for entertainment were less likely to recall candidates, less likely to know political facts, and less likely to vote, *only when they had access to new media*. Cable and Internet access, in other words, provide people who want to avoid politics with the means to do so. According to Baum (1999, 18), "the highly segmented modern television marketplace presumably allows individuals to escape political news more effectively than was the case in prior decades." This study suggests that escaping politics is made even easier with access to the Internet. While it would be premature to conclude that the impact of cable and Internet is additive in this respect, the analysis does suggest that access to both new media leads to a stronger interaction with entertainment preference than access to any one of them.

In their study of the effects of different media environments, Delli Carpini et al. (1994, 454) emphasize that "learning requires not only the will to learn, but the opportunity to do so." For all practical purposes, people with access to new media have endless opportunity to learn about politics. But at the same time, people's "will to learn" becomes much more important. The results of this study suggest that for some people, political learning in the broadcast era was not driven by the "will to learn", but simply by the lack of other options to satisfy their desire to be entertained. Knowledge and vote likelihood of people with a high entertainment preference was lower for new media users across different data sets and operationalizations of the concept of entertainment preference. To turn Delli Carpini et al.'s above statement on its head, avoiding news requires not only the will to watch something else, but also the opportunity to do so.

Support for the other side of the interaction effect is more ambiguous. People who prefer news and have new media access do not appear to know more about politics than similar respondents without this access and the chance to watch or read news at any time. This result does not seem to be explained by a ceiling effect. It is all the more puzzling since the likelihood to vote among people with a news preference is indeed higher for new media users. Future research will have to resolve the details of this argument. At least with respect to turnout, results clearly suggest that the gap between people who prefer entertainment and people who like news better is larger, the more access to new media people have.

Content preferences are better predictors of political knowledge and turnout among new media users in the present cross-sectional analysis. It is tempting to infer that the growth of new media is the cause of lower news audiences and declining knowledge and turnout among people who prefer entertainment programming. Direct evidence for this could only come from panel data covering people's transition from recipients of over-the-air television to cable and Internet users. But the likely chain of events is obvious: Most people do not watch television to follow a particular program. They decide to watch and then turn to the best available program. During the broadcast era, simultaneous broadcast of news on the three networks led most people to watch some news, even those who were not particularly interested in politics and would have turned to entertainment in a heartbeat, had that been possible. In Neuman's (1996, 19) words, people watched "politics by default". Cable television and the Internet removed these constraints and enabled subscribers to better match content preferences and content. News audiences have not declined over the course of a decade or two because people became dramatically less interested in news and politics. While that may be part of the explanation, the present study suggests a different cause. A segment of the electorate was never particularly interesting and watched news merely out of habit and for lack of better options. When cable subscribers and Internet users were offered greater content choice, the consequence was a decline in knowledge and turnout among people who only watched news because they were "captive". This argument leads to the uncomforting conclusion that even the mediocre levels of political knowledge during the broadcast era (e.g., Delli Carpini & Keeter, 1996) were, in part, a result of de facto restrictions of people's freedom to choose their preferred media content.

# Appendix

*NES Variables*

Note: All calculations in the paper use weights. Weights for NES 2000 are in v000002, for NES 1996 in v960003.

**Cable:** 1 if respondent subscribes to cable or satellite television (v000334, v960241).
**Internet Access:** v961160, v001433.
**Party ID:** v000523, v960420
**Incumbent Candidate ID:** 1 if respondent identifies with the party of the congressional incumbent candidate, 0 else.
**Open Race:** 1 if no incumbent candidate runs in congressional election.
**Strong Party Identifier/Weak Party Identifier/Independent Leaner:** Coded from party ID.
**Education:** v000913, v960610
**Income:** v000994, v960701
**Gender:** v001029, v960066
**Race (Black, Hispanic, other Non-White):** Dummy variables based on v001006a, v960067. Format for NES 2000 was changed to allow self-classification as Hispanic.
**Age:** v000908, v960605
**South:** 1 if respondent lives in Alabama, Arkansas, Florida, Georgia, Louisiana, Mississippi, North Carolina, South Carolina, Tennessee, Texas, or Virginia, 0 otherwise (v000079, v960109).
**Weekly Working Hours :** Number of hours respondent works per week (v000985, v960670): 1-20:1, 21-30:2, 31-40:3, 41-50:4, 51-60:5, 61-70:6, and 71-97:7 (2000 uses continuous scale). Respondents who reported to be unemployed, retired without other occupation or permanently disabled and not working in v000919 (v960615) are coded as 0.
**Religious attendance:** Five-point scale from "never" to "every week" (v000877, v000879, v960576, v960578).
**Campaign Interest:** v001201, v960201
**Political Information:** Interviewer rating of respondents political information (v001745, v960070).
**Discuss Politics:** Frequency of discussion of politics with friends or family (v001205, v961005).
**Group Memberships:** NES 2000: Total number of groups in which respondent is a member (v001495). NES 1996: Total number of groups in which respondent is member and reports political discussion "often" or "sometimes" (v961458).
**Trust:** Respondents were assigned a value of 1 for answering "Just about always" or "Most of the time" on "How much of the time do you think you can trust the government in Washington to do what is right?" (v001534, v961251); for "Waste some" or "Don't waste very much" on "Do you think that people in government waste a lot of the money we pay in taxes?" (v001535, v961252); for "Gov't run by a few big interests" on "Would you say the government is pretty much run by a few big interests looking out for themselves or that it is run for the benefit of all the people?" (v001536, v961253); and for "Quite a few are crooked" or "Not very many are crooked" on "Do you think that quite a few of the people running the government are crooked, not very many are, or do you think hardly any of them are crooked?" (v001537, v961254). The trust scale is the average on these four items.
**Internal Efficacy:** Degree of agreement with the statement that "Sometimes politics and government seem so complicated that a person like me can't really understand what's going on." (v001529, v961246)
**External Efficacy:** Respondents were assigned a value of 1 if they disagree somewhat or strongly with the statement that "Public officials don't care much what people like me think" (v001527, v961244). Neither agree or disagree and don't know were coded as .5, other answers as 0. The same values were assigned for agreement that "People like me don't have any say about what the government does" (v001528, v961245). The external efficacy index is the average on the two items.
**Live in Community:** 0 if respondents lives in present community for less than a year, 2 if at least one, but less than 2 years, 2 else (v001020c, v960712).
**Watch National News:** Number of days in past week respondent watched national news (transformed to 0-1 interval) (v000329, v960242).
**Watch Local News:** Number of days in past week respondent watched local news (v960244).
**Watch Early/Late Local News:** Number of days in past week respondent watched early/ late local news (transformed to 0-1 interval) (v000331, v000332).
**Read Newspaper:** Number of days in past week respondent read a newspaper (v000335, v960246).
**Listen to Talk Radio:** 1 if respondent listens to talk radio, 0 otherwise (v001430, v961155).
**Time of Interview:** Days between interview and election (v000131, v960904).

*Pew Media Consumption Surveys*

Note: All calculations in the paper use weights. Since weights do not sum to zero, unweighted N's are reported.

**Cable:** Respondent subscribes to cable TV (1996: Question 38A, 1998: 67, 2000: 33).
**Internet Access:** 1996: "Do you ever use a computer at work, school or home to connect with computer bulletin boards, information services such as America Online or Prodigy, or other computers over the Internet? (Q. 36); 1998: "Do you ever use a computer at work, school or home to connect with computers over the Internet, the World Wide Web, or with information services such as America Online or Prodigy?" (Q. 56); 2000: "Do you ever go online to access the Internet or World Wide Web or to send and receive email?" (Q. 38)
**Party ID:** Five-point scale: Republican, leaning Republican, independent/no preference, leaning Democratic, Democratic (1996: D.10, D.11, 1998: D.11, D.12, 2000: D.11, D.12).
**Strong Party Identifier/Weak Party Identifier:** Coded from party ID.
**Education:** 1996: EDUC, 1998: D.5, 2000: D.5
**Income:** 1996: INCOME, 1998: D.10, 2000: D.10
**Gender:** 1996: D.1, 1998: D.1, 2000: D.1
**Age:** 1996: AGE, 1998: D.2, 2000: D.2
**Race (Black, Asian, other Non-White):** 1996: RACE, 1998: D.7, 2000: D.7
**Owns Home:** 1 if respondent own her home, 0 otherwise (1996: HOME, 1998: D.17, 2000: D.17).
**Marital Status:** 1996: D.3, 1998: D.3, 2000: D.3
**Has Children under 18:** Respondent has children under 18 living in her household (1996: D.3A, 1998: D.4A, 2000: D.4A)
**Employed:** 1 if respondent works full-time, 0 otherwise (1996: D.19A, 1998: D.18, 2000: D.18)
**Size of Town:** Size of the place where respondent lives (four-point scale) (1996: D.21, 1998: D.19, 2000: D.20)
**Watch National News:** Frequency of watching (four-point scale) (1996: 13A, 1998:16A, 2000: 16A)
**Watch Local News:** Frequency of watching (four-point scale) (1996: 13B, 1998: 16B, 2000: 16B)
**Read Newspaper:** 1 if respondents reads one or more daily newspapers "regularly", 0 otherwise (1996: 2, 1998: 3, 2000: 4)

**Listen to Talk Radio:** Frequency of listening to political call-in shows (four-point scale) (1998: 4, 1998: 31A, 2000: 19).
**Reads Magazines:** Frequency of reading (four-point scale) (1996: 14A, 1998: 18A, 2000: 17A)
**Follows News:** 1996: How closely did respondent follow stories about Bosnia (6A), Republican presidential candidates (6B), death of Ron Brown (6C), Unabomber case (6D), Clinton's veto of abortion bill (6E), Middle East (6F), law on domestic terrorism (6G), and plan crash (6H). Index is average on these eight four-point scales.
1998: How closely did respondent follow coverage of allegations against Clinton (17A), regulation of tobacco industry (17B), peace agreement between Great Britain and Ireland (17D), and "candidates and election campaigns in your state" (17E). Index is average on these four four-point scales.
2000: How closely did respondent follow stories about candidates for the 2000 presidential election (8A), and about the U.S. stock market (8B). Index is average on these two four-point scales.

Table 1: Candidate Name Recall (NES 1996, 2000)

|  | NES 1996 | | NES 2000 | |
| --- | --- | --- | --- | --- |
|  | *Model 1* | *Model 2* | *Model 1* | *Model 2* |
| Entertainment Preference | .30 (.15)** | .17 (.14) | .11 (.12) | -.021 (.067) |
| Cable | .052 (.065) | -- | -.024 (.045) | -- |
| New Media | -- | .0039 (.046) | -- | -.025 (.025) |
| **Ent. Pref. X Cable** | **-.47 (.18)**** | -- | **-.19 (.13)** | -- |
| **Ent. Pref. X New Media** | -- | **-.21 (.13)*** | -- | **-.12 (.060)*** |
| Party ID | .0092 (.011) | .0090 (.011) | -.0080 (.0093) | -.0084 (.0093) |
| Incumbent Cand. ID | -.039 (.025) | -.039 (.025) | .010 (.018) | .011 (.018) |
| Open Race | -.0031 (.074) | -.0043 (.074) | .095 (.060) | .094 (.059) |
| Strong Party Identifier | .19 (.088)** | .19 (.088)** | .081 (.055) | .077 (.055) |
| Weak Party Identifier | .13 (.084) | .13 (.084) | .010 (.048) | .0079 (.049) |
| Independent Leaner | .12 (.087) | .12 (.072) | .016 (.049) | .016 (.049) |
| Education | .064 (.017)*** | .067 (.017)*** | .041 (.013)*** | .042 (.013)*** |
| Income | .013 (.0042)*** | .013 (.0043)*** | .0088 (.0060) | .0087 (.0058) |
| Gender | -.054 (.046) | -.06 (.046) | .0034 (.033) | .0043 (.033) |
| Black | -.42 (.066)*** | -.43 (.067)*** | -.15 (.054)*** | -.14 (.053)*** |
| Hispanic | -- | -- | -.11 (.053)** | -.12 (.054)** |
| Other Non-White | -.31 (.15)** | -.30 (.15)** | -.0073 (.067) | -.0021 (.068) |
| Age | .019 (.0073)*** | .019 (.0074)** | .0032 (.0050) | .0026 (.0051) |
| Age$^2$ | -.0002 (.0001)*** | -.0002 (.0001)*** | -.00002 (.00005) | -.00002 (.00005) |
| South | -.073 (.045) | -.068 (.048) | -.027 (.034) | -.025 (.034) |
| Weekly Working Hours | -.002 (.0012) | -.0020 (.0013) | -.0018 (.00085)** | -.0019 (.00086)** |
| Religious Attendance | .13 (.058)** | .12 (.059)** | .096 (.043)** | .10 (.044)** |
| Political Information | .49 (.11)*** | .47 (.12)*** | .35 (.074)*** | .37 (.075)*** |
| Campaign Interest | .17 (.073)** | .18 (.073)** | .022 (.054) | .027 (.054) |
| Discuss Politics | -.07 (.085) | -.062 (.086) | .049 (.043) | .048 (.043) |
| Group Memberships | .018 (.016) | .019 (.016) | .036 (.013)*** | .037 (.013)*** |
| Trust | .049 (.069) | .063 (.070) | -.0067 (.049) | -.0064 (.049) |
| Internal Efficacy | .086 (.079) | .081 (.080) | .082 (.054) | .090 (.055)* |
| External Efficacy | .089 (.063) | .095 (.063) | -.037 (.045) | -.035 (.045) |
| Live in Community | .22 (.055)*** | .22 (.055)*** | .099 (.025)*** | .096 (.025)*** |
| Watch National News | -.017 (.072) | -.0090 (.073) | .014 (.051) | .010 (.051) |
| Watch (Early/Late) Local News | .022 (.068) | .027 (.069) | .027 (.050) / -.035 (.047) | .021 (.050) / -.035 (.048) |
| Read Newspaper | .14 (.057)** | .13 (.057)** | .013 (.0058)** | .013 (.0059)** |
| Listen to Talk Radio | -.0076 (.048) | -0044 (.048) | .032 (.037) | .032 (.037) |
| Time of Interview | -.0078 (.0027)*** | -.0078 (.0027)*** | -.0044 (.0014)*** | -.0044 (.0014)*** |
| Constant | -.96 (.22)*** | -.92 (.22)*** | -.39 (.14)*** | -.35 (.14)** |
| N (with MV imputations) | 1448 | 1441 | 1358 | 1347 |
| N (without MV imputations) | 1260 | 1259 | 1083 | 1078 |
| R$^2$ | .24 | .23 | .22 | .22 |

*** p<.01, ** p<.05, * p<.10

Note: Models are OLS regressions estimated on five imputed data sets created by AMELIA (Honaker et al., 1999). The coefficient estimate is the mean of the five separate estimates. The robust standard error (in parentheses) is based on the variance across the five imputed data sets plus the variance within each data set. For details, see King et al. (2001).
The table lists the number of observations with and without multiple imputations of missing values. Goodness-of-fit statistics are only available for non-imputed estimations.

Table 2: Knowledge of Incumbent House Candidates (NES 1996, 2000)

|  | NES 1996 | | NES 2000 | |
| --- | --- | --- | --- | --- |
|  | *Model 1* | *Model 2* | *Model 1* | *Model 2* |
| Entertainment Preference | .15 (.056)*** | .14 (.053)*** | .0083 (.061) | .041 (.041) |
| Cable | .029 (.024) | -- | -.039 (.023)* | -- |
| New Media | -- | .023 (.016) | -- | -.029 (.013)** |
| **Ent. Pref. X Cable** | **-.13 (.066)**** | -- | **-.0035 (.069)** | -- |
| **Ent. Pref. X New Media** | -- | **-.094 (.045)**** | -- | **-.098 (.046)**** |
| Party ID | -.0025 (.0036) | -.0021 (.0037) | -.0072 (.0043)* | -.0077 (.0043)* |
| Incumbent Cand. ID | .0088 (.0080) | .0085 (.0080) | .020 (.0090)** | .020 (.0089)** |
| Strong Party Identifier | .077 (.029)*** | .075 (.029)*** | .042 (.026) | .039 (.026) |
| Weak Party Identifier | .070 (.027)** | .069 (.027)** | .028 (.026) | .025 (.026) |
| Independent Leaner | .087 (.029)*** | .088 (.028)*** | .045 (.026)* | .043 (.026)* |
| Education | .0032 (.0059) | .0028 (.0059) | .017 (.0061)*** | .018 (.0062)*** |
| Income | -.0017 (.0016) | -.0022 (.0017) | .0030 (.0025) | .0046 (.0028) |
| Gender | -.041 (.015)*** | -.042 (.015)*** | -.021 (.017) | -.018 (.017) |
| Black | -.058 (.030)* | -.059 (.030)** | -.094 (.028)*** | -.098 (.027)*** |
| Hispanic | -- | -- | -.082 (.040)** | -.084 (.040)** |
| Other Non-White | -.039 (.059) | -.039 (.060) | -.046 (.034) | -.040 (.034) |
| Age | .012 (.0027)*** | .012 (.0027)*** | .011 (.0028)*** | .010 (.0028)*** |
| Age$^2$ | -.0001 (.00003)*** | -.0001 (.00003)*** | -.0001 (.00003)*** | -.0001 (.00003)*** |
| South | .0053 (.016) | .0063 (.016) | -.039 (.018)** | -.040 (.018)** |
| Weekly Working Hours | -.00068 (.00042) | -.00069 (.00042)* | -.00046 (.00043) | -.00045 (.00043) |
| Religious Attendance | .022 (.019) | .023 (.019) | .0091 (.021) | .012 (.021) |
| Political Information | .24 (.036)*** | .23 (.036)*** | .23 (.038)*** | .24 (.038)*** |
| Campaign Interest | .057 (.025)** | .060 (.025)** | .046 (.029) | .048 (.029)* |
| Discuss Politics | .047 (.029) | .045 (.029) | .030 (.022) | .033 (.022) |
| Group Memberships | .0084 (.0050)* | .0088 (.0050)* | .013 (.0048)*** | .013 (.0048)*** |
| Trust | .015 (.023) | .017 (.023) | -.0013 (.025) | .0027 (.025) |
| Internal Efficacy | .042 (.027) | .043 (.027) | .040 (.026) | .048 (.026)* |
| External Efficacy | .011 (.021) | .011 (.021) | .024 (.022) | .021 (.022) |
| Live in Community | .094 (.022)*** | .093 (.022)*** | .063 (.019)*** | .057 (.020)*** |
| Watch National News | .013 (.024) | .011 (.024) | .013 (.026) | .0081 (.026) |
| Watch (Early/Late) Local News | .053 (.024)** | .056 (.024)** | .049 (.024)** / .020 (.022) | .048 (.024)*/ .016 (.022) |
| Read Newspaper | .046 (.020)** | .044 (.020)** | .012 (.0032)*** | .012 (.0032)*** |
| Listen to Talk Radio | .027 (.016)* | .026 (.016) | .013 (.017) | .013 (.017) |
| Time of Interview | -.0001 (.00093) | -.00016 (.00094) | -.0003 (.00076) | -.00021 (.00075) |
| Constant | -.20 (.084)** | -.20 (.084)** | -.23 (.080)*** | -.22 (.079)*** |
| N (with MV imputations) | 1315 | 1309 | 1352 | 1341 |
| N (without MV imputations) | 1135 | 1136 | 1083 | 1076 |
| R$^2$ | .22 | .24 | .26 | .33 |

\*\*\* *p*<.01, \*\* *p*<.05, \* *p*<.10

Note: Models are OLS regressions estimated on five imputed data sets created by AMELIA (Honaker et al., 1999). The coefficient estimate is the mean of the five separate estimates. The robust standard error (in parentheses) is based on the variance across the five imputed data sets plus the variance within each data set. For details, see King et al. (2001).
The table lists the number of observations with and without multiple imputations of missing values. Goodness-of-fit statistics are only available for non-imputed estimations.

Table 3: Voting in 1996 Presidential and House Elections (NES 1996)

|  | House Election | | Presidential Election | |
|---|---|---|---|---|
|  | *Model 1* | *Model 2* | *Model 1* | *Model 2* |
| Entertainment Preference | 1.63 (.56)*** | 1.45 (.55)*** | .99 (.61) | .74 (.61) |
| Cable | .70 (.24)*** | -- | .26 (.27) | -- |
| New Media | -- | .47 (.18)** | -- | .22 (.20) |
| **Ent. Pref. X Cable** | **-1.47 (.66)** | -- | **-.36 (.74)** | -- |
| **Ent. Pref. X New Media** | -- | **-.92 (.49)*** | -- | **.0055 (.58)** |
| Party ID | .058 (.045) | .056 (.045) | -.017 (.049) | -.016 (.048) |
| Incumbent Party ID | .088 (.089) | .098 (.089) | -.065 (.097) | -.061 (.096) |
| Open Race | .32 (.24) | .31 (.24) | -.14 (.25) | -.10 (.25) |
| Strong Party Identifier | 1.65 (.31)*** | 1.60 (.31)*** | 1.77 (.32)*** | 1.72 (.33)*** |
| Weak Party Identifier | .83 (.28)*** | .80 (.28)*** | .87 (.29)*** | .86 (.27)*** |
| Independent Leaner | .58 (.30)* | .58 (.30)* | .77 (.31)** | .75 (.31)** |
| Education | .18 (.065)*** | .16 (.066)** | .27 (.073)*** | .26 (.074)*** |
| Income | .051 (.016)*** | .051 (.016)*** | .066 (.017)*** | .063 (.017)*** |
| Gender | .31 (.17)* | .30 (.17)* | .19 (.18) | .19 (.18) |
| Black | -.32 (.26) | -.30 (.26) | -.28 (.26) | -.25 (.26) |
| Other Non-White | -.69 (.50) | -.76 (.51) | -.48 (.62) | -.46 (.62) |
| Age | .060 (.028)** | .069 (.028)** | .021 (.031) | .026 (.031) |
| Age$^2$ | -.00043 (.00027) | -.00051 (.00027)* | -.000066 (.00030) | -.00011 (.00030) |
| South | -.25 (.17) | -.26 (.17) | -.43 (.18)** | -.44 (.18)** |
| Weekly Working Hours | .0012 (.0049) | .00092 (.0048) | -.0065 (.0054) | -.0062 (.0055) |
| Religious Attendance | 1.12 (.22)*** | 1.11 (.23)*** | 1.13 (.24)*** | 1.10 (.24)*** |
| Political Information | 1.52 (.41)*** | 1.46 (.41)*** | 1.79 (.43)*** | 1.70 (.44)*** |
| Campaign Interest | .76 (.27)*** | .80 (.27)*** | 1.30 (.29)*** | 1.33 (.29)*** |
| Discuss Politics | .87 (.32)*** | .84 (.33)** | .76 (.34)** | .74 (.35)** |
| Group Memberships | .10 (.079) | .096 (.077) | .082 (.099) | .080 (.098) |
| Trust | -.091 (.25) | -.039 (.25) | -.12 (.26) | -.098 (.26) |
| Internal Efficacy | -.012 (.30) | .0091 (.30) | -.18 (.32) | -.20 (.32) |
| External Efficacy | .41 (.23)* | .36 (.23) | .47 (.26)* | .46 (.26)* |
| Live in Community | .75 (.23)*** | .73 (.23)*** | .36 (.21)* | .38 (.21)* |
| Watch National News | .62 (.25)** | .60 (.25)** | .28 (.26) | .26 (.27) |
| Watch Local News | .086 (.27) | .12 (.26) | -.16 (.27) | -.14 (.27) |
| Read Newspaper | .16 (.21) | .16 (.21) | .33 (.22) | .33 (.22) |
| Listen to Talk Radio | .042 (.18) | .045 (.18) | .16 (.19) | .17 (.19) |
| Time of Interview | -.0039 (.0094) | -.0047 (.0096) | .0043 (.010) | .0041 (.010) |
| Constant | -7.98 (.98)*** | -7.98 (1.00)*** | -5.86 (.95)*** | -5.92 (.97)*** |
| N (with MV imputations) | 1437 | 1430 | 1448 | 1441 |
| N (without MV imputations) | 1260 | 1249 | 1250 | 1259 |
| Pseudo R$^2$ | .32 | .30 | .30 | .32 |

\*\*\* *p*<.01, \*\* *p*<.05, \* *p*<.10

Note: Models are logit regressions estimated on five imputed data sets created by AMELIA (Honaker et al., 1999). The coefficient estimate is the mean of the five separate estimates. The robust standard error (in parentheses) is based on the variance across the five imputed data sets plus the variance within each data set. For details, see King et al. (2001).
The table lists the number of observations with and without multiple imputations of missing values. Goodness-of-fit statistics are only available for non-imputed estimations.

Table 4: Voting in 1992 Presidential and 1994 House Elections (Pew MCS 1996)

|  | Voted in 1992 Pres. Election | | Vote in 1994 House Election | |
|---|---|---|---|---|
|  | *Model 1* | *Model 2* | *Model 1* | *Model 2* |
| Entertainment Pref. | -.72 (.61) | -.71 (.59) | .20 (.66) | -.27 (.65) |
| Cable | .62 (.25)* | -- | .55 (.25)** | -- |
| New Media | -- | .56 (.19)*** | -- | .40 (.17)** |
| **Ent. Pref. X Cable** | **-1.16 (.74)** | -- | **-1.91 (.80)**** | -- |
| **Ent. Pref. X New Media** | -- | **-.88 (.52)*** | -- | **-.96 (.54)*** |
| Party ID | .095 (.040)** | .095 (.040)** | -.072 (.039)* | -.075 (.039)* |
| Strong Party Identifier | 1.27 (.21)*** | 1.25 (.21)*** | .92 (.23)*** | .88 (.23)*** |
| Weak Party Identifier | .87 (.23)*** | .84 (.22)*** | .76 (.24)*** | .74 (.24)*** |
| Education | .22 (.045)*** | .22 (.045)*** | .21 (.045)*** | .21 (.045)*** |
| Income | .035 (.038) | .028 (.038) | .089 (.041)** | .083 (.041)* |
| Gender | -.31 (.12)** | -.31 (.13)** | -.018 (.130) | -.017 (.13) |
| Age | .10 (.021)*** | .11 (.021)*** | .10 (.025)*** | .10 (.025)*** |
| Age$^2$ | -.0007 (.0002)*** | -.0008 (.0002)*** | -.0006 (.0003)** | -.0006 (.0003)** |
| Black | -.13 (.21) | -.13 (.21) | -.17 (.24) | -.18 (.23) |
| Other Non-White | -.70 (.24)*** | -.68 (.24)*** | -.69 (.26)*** | -.66 (.26)** |
| Owns Home | .21 (.15) | .21 (.15) | .20 (.16) | .19 (.16) |
| Employment Status | .20 (.14) | .20 (.14) | -.35 (.14)** | -.36 (.14)*** |
| Size of Town | -.016 (.060) | -.010 (.060) | .020 (.063) | .028 (.063) |
| Watch National News | -.067 (.21) | -.089 (.21) | -.26 (.22) | -.26 (.22) |
| Watch Local News | .058 (.26) | .093 (.26) | .34 (.29) | .36 (.29) |
| Read Newspaper | .13 (.15) | .12 (.15) | .31 (.16)** | .30 (.16)* |
| Listen to Talk Radio | .31 (.18)* | .30 (.18)* | .61 (.19)*** | .60 (.19)*** |
| Reads Magazines | .042 (.19) | .011 (.19) | .21 (.20) | .18 (.20) |
| Follows News | .71 (.36)** | .72 (.36)** | 1.11 (.37)*** | 1.14 (.37)*** |
| Constant | -4.91 (.63)*** | -5.12 (.63)*** | -6.34 (.71)*** | -6.39 (.71)*** |
| N (with MV imputations) | 1708 | 1708 | 1585 | 1585 |
| N (without MV imputations) | 1521 | 1521 | 1424 | 1424 |
| Pseudo R$^2$ | .16 | .16 | .20 | .20 |

*** *p*<.01, ** *p*<.05, * *p*<.10

Note: Models are logit regressions estimated on five imputed data sets created by AMELIA (Honaker et al., 1999). The coefficient estimate is the mean of the five separate estimates. The robust standard error (in parentheses) is based on the variance across the five imputed data sets plus the variance within each data set. For details, see King et al. (2001).
The table lists the number of observations with and without multiple imputations of missing values. Goodness-of-fit statistics are only available for non-imputed estimations.

Table 5: Voting in 1996 Presidential Election (Pew MCS 1998, 2000)

|  | Pew MCS 1998 | | Pew MCS 2000 | |
| --- | --- | --- | --- | --- |
|  | *Model 1* | *Model 2* | *Model 1* | *Model 2* |
| Entertainment Pref. | .12 (.41) | .20 (.42) | .20 (.32) | .62 (.39) |
| Cable | .17 (.15) | -- | .43 (.16)*** | -- |
| New Media | -- | .28 (.11)*** | -- | .48 (.11)*** |
| **Ent. Pref. X Cable** | **-.68 (.48)** | -- | **-1.03 (.40)**\*\* | -- |
| **Ent. Pref. X New Media** | -- | **-.57 (.32)**\* | -- | **-.91 (.27)**\*\*\* |
| Party ID | .096 (.029)*** | .097 (.029)*** | -.038 (.033) | -.040 (.033) |
| Strong Party Identifier | 1.53 (.14)*** | 1.52 (.14)*** | 1.10 (.14)*** | 1.09 (.14)*** |
| Weak Party Identifier | .80 (.16)*** | .80 (.16)*** | .45 (.16)*** | .45 (.16)*** |
| Education | .22 (.033)*** | .21 (.034)*** | .32 (.038)*** | .30 (.039)*** |
| Income | .054 (.028)* | .044 (.029) | .037 (.039) | .025 (.040) |
| Gender | .0041 (.094) | -.0017 (.094) | .053 (.11) | .046 (.11) |
| Age | .066 (.015)*** | .067 (.015)*** | .11 (.015)*** | .11 (.016)*** |
| Age$^2$ | -.0004 (.0002)*** | -.0004 (.0002)*** | -.0007 (.0002)*** | -.0007 (.0002)*** |
| Black | .032 (.14) | .038 (.15) | .64 (.19)*** | .66 (.19)*** |
| Other Non-White | -.22 (.24) | -.22 (.24) | -.51 (.21)** | -.53 (.21)** |
| Owns Home | .39 (.11)*** | .41 (.11)*** | .43 (.12)*** | .44 (.12)*** |
| Employment Status | .18 (.11) | .17 (.11) | .16 (.12) | .17 (.12) |
| Size of Town | -.072 (.045) | -.064 (.045) | -.020 (.051) | -.011 (.051) |
| Watch National News | .13 (.14) | .13 (.14) | .17 (.15) | .19 (.15) |
| Watch Local News | -.071 (.18) | -.087 (.18) | .059 (.19) | .060 (.19) |
| Read Newspaper | .36 (.11)*** | .35 (.11)*** | .22 (.11)** | .22 (.11)** |
| Listen to Talk Radio | -.048 (.094) | -.047 (.094) | .22 (.12)* | .21 (.12)* |
| Reads Magazines | .20 (.14) | .19 (.14) | .078 (.16) | .040 (.16) |
| Follows News | 1.01 (.24)*** | 1.00 (.24)*** | 1.09 (.21)*** | 1.05 (.21)*** |
| Constant | -5.30 (.45)*** | -5.40 (.45)*** | -6.27 (.48)*** | -6.53 (.49)*** |
| N (with MV imputations) | 2918 | 2918 | 2736 | 2736 |
| N (without MV imputations) | 2426 | 2426 | 2260 | 2260 |
| Pseudo R$^2$ | .18 | .19 | .24 | .24 |

\*\*\* *p*<.01, \*\* *p*<.05, \* *p*<.10

Note: Models are logit regressions estimated on five imputed data sets created by AMELIA (Honaker et al., 1999). The coefficient estimate is the mean of the five separate estimates. The robust standard error (in parentheses) is based on the variance across the five imputed data sets plus the variance within each data set. For details, see King et al. (2001).
The table lists the number of observations with and without multiple imputations of missing values. Goodness-of-fit statistics are only available for non-imputed estimations.

Table A.1: Measuring Relative Entertainment Preference

| | Items used to measure… | | | | | Percentiles | | | Correlation with… | | |
|---|---|---|---|---|---|---|---|---|---|---|---|
| | Entertainment Viewing | Information Viewing | N | Mean | Stand. Dev. | 25th | 50th | 75th | Campaign Interest | Attention to News | Follow Ent. News |
| NES 2000 | *Jeopardy*, *Wheel of Fortune* | Early Local News, National News | 1356 | .18 | .26 | .00 | .00 | .31 | -.13 | -.24[1] / -.13[2] | |
| NES 1996 | *Jeopardy* or *Wheel of Fortune*, *Dr. Quinn* | Local News, National News | 1446 | .21 | .24 | .00 | .18 | .35 | -.12 | -.22[1] / -.17[2] | |
| Pew MCS 2000 | *Entertainment Tonight* | Local News, National News | 2964 | .28 | .26 | .00 | .28 | .50 | | -.09[3] | .30 |
| Pew MCS 1998 | *Entertainment Tonight*, MTV | Local News, National News | 2918 | .25 | .21 | .00 | .25 | .40 | | -.09[3] | .30 |
| Pew MCS 1996 | *Hardcopy*, MTV | Local News, National News | 1708 | .30 | .19 | .17 | .33 | .40 | | -.07[3] | .18 |

Note: Number of cases are based on unweighted samples. Means and standard deviations are calculated using weights.

[1] Attention to national news
[2] Attention to local news
[3] Composite index of following various news stories (see appendix for specific items)

Table A.2: Relative Entertainment Preference and Exposure to News

|  | Pew MCS 1996 | Pew MCS 1996 | Pew MCS 2000 | NES 1996 | NES 2000 |
|---|---|---|---|---|---|
|  | *Time Watching News* | *More news than before* | *Watch News Regularly*[2] | *News Exposure* | *News Exposure* |
| Entertainment Preference | .054 (.10) | .47 (.19)** | .40 (.35) | .0064 (.065) | .037 (.078) |
| Cable | .14 (.043)*** | .23 (.062)*** | .36 (.15)** | -.0034 (.025) | .10 (.028)*** |
| **Ent. Pref. X Cable** | **-.35 (.13)*** | **-.70 (.21)*** | **-.74 (.43)* | **-.098 (.074)** | **-.15 (.088)* |
| Education | .00068 (.0072) | -.0085 (.0092) | -.048 (.030) | .035 (.0053)*** | .040 (.0061)*** |
| Employment Status[1] | -.085 (.023)*** | -.045 (.029) | -.42 (.098)*** | -.0022 (.00038)*** | -.0013 (.00042)*** |
| Black | .099 (.040)** | .062 (.046) | .58 (.19)*** | .0027 (.026) | -.081 (.027)*** |
| Hispanic | -- | -- | -- | -- | -.030 (.039) |
| Other Non-White | -.011 (.040) | -.052 (.057) | -.068 (.19) | -.020 (.044) | -.025 (.040) |
| Constant | .48 (.047)*** | .55 (.064)*** | 1.43 (.18)*** | .35 (.031)*** | .36 (.039)*** |
| Adj. $R^2$ | .0024 | .035 | .016 | .057 | .07 |
| N | 1679 | 829 | 2903 | 1431 | 1336 |

\*\*\* $p<.01$, \*\* $p<.05$, \* $p<.10$

Note: Models are estimated by OLS regression. Cell entries are unstandardized coefficient estimates and robust standard errors in parentheses. All models are calculated using sample weights.

[1] Weekly working hours for NES 1996 and 2000.
[2] Logit regression since dependent variable is dichotomous.

Figure 1: Political Knowledge

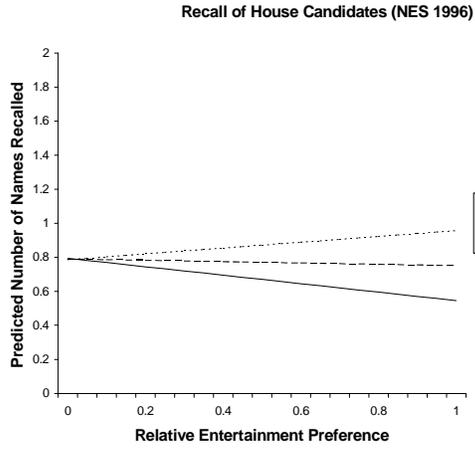
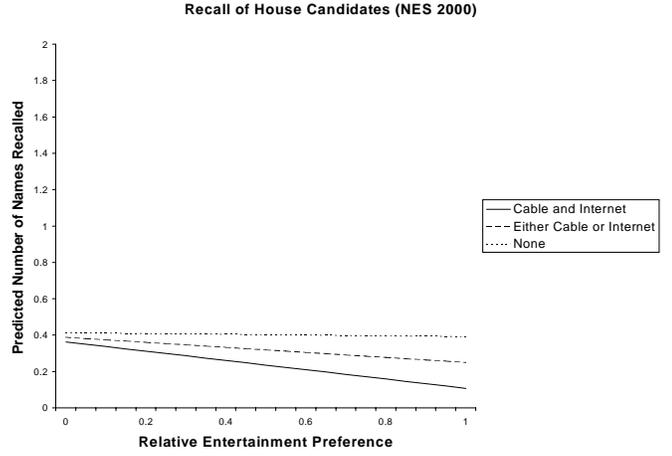
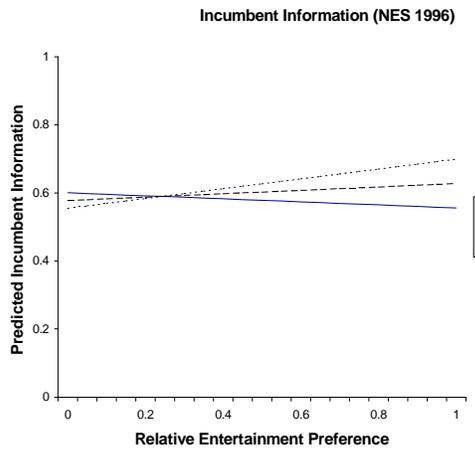
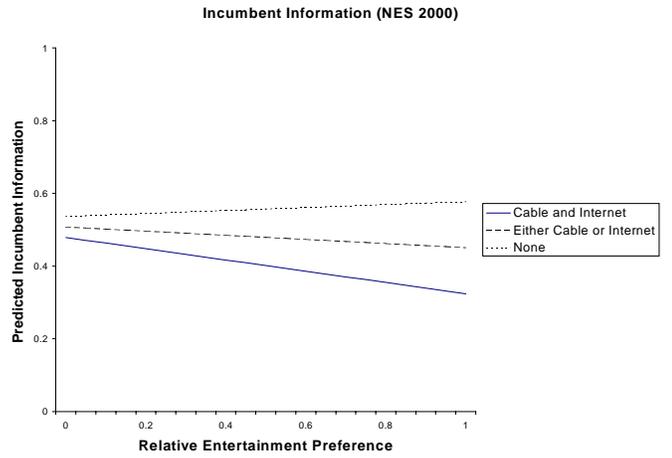

Figure 2: Vote Likelihood

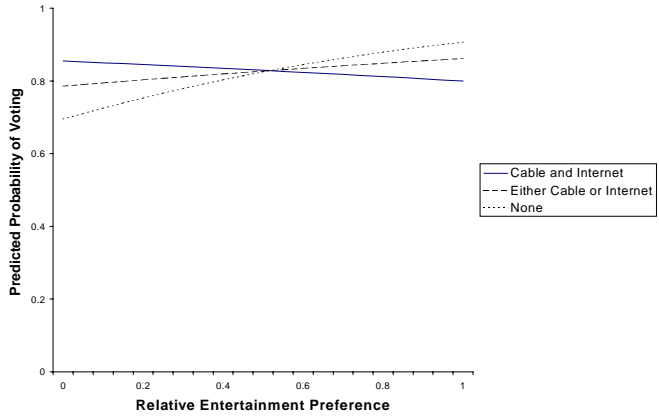

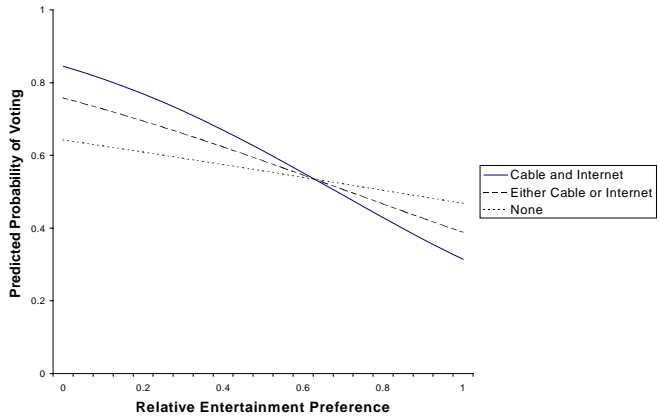

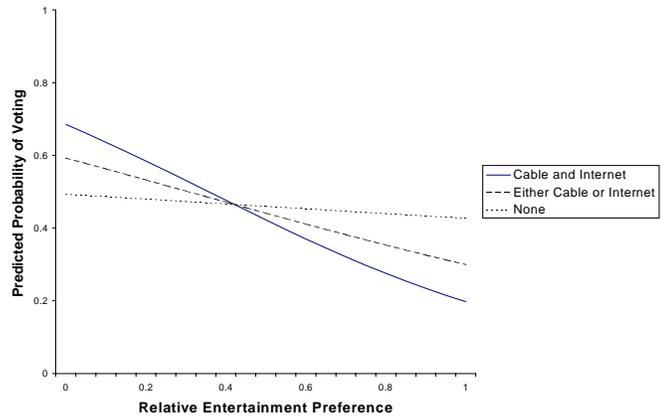

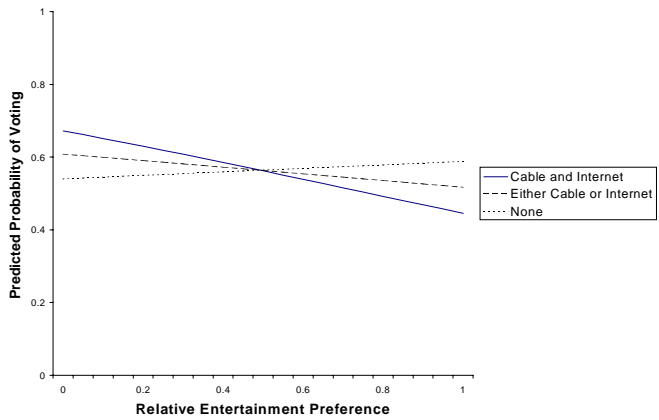

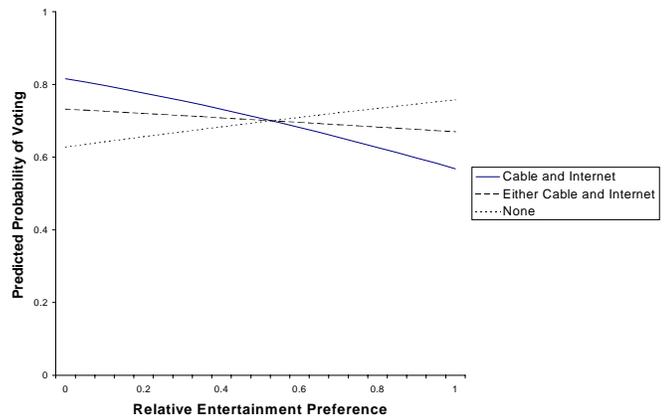